\documentclass{article}
\usepackage[utf8]{inputenc}

\usepackage{amsthm}
\newtheorem{theorem}{Theorem}[section]
\newtheorem{corollary}[theorem]{Corollary}
\newtheorem{lemma}[theorem]{Lemma}
\newtheorem{claim}[theorem]{Claim}
\newtheorem{conjecture}[theorem]{Conjecture}

\usepackage{graphicx}
\usepackage{comment}
\usepackage{hyperref}

\title{Parameterized complexity of \\ {\sc Bandwidth} of Caterpillars \\
and {\sc Weighted Path Emulation}}
\author{Hans L. Bodlaender\thanks{Department of Information and Computing Sciences, Utrecht University, P.O. Box 80.089, 3508 TB Utrecht, the Netherlands. Email: \href{mailto:h.l.bodlaender@uu.nl}{h.l.bodlaender@uu.nl}}}
\date{}

\begin{document}

\maketitle 

\begin{abstract}
    In this paper, we show that {\sc Bandwidth} is hard for the complexity class $W[t]$ for all $t\in {\bf N}$, even for caterpillars with hair length at most three.
    As intermediate problem, we introduce the {\sc Weighted Path Emulation} problem: given
    a vertex-weighted path $P_N$ and integer $M$,
    decide if there exists a
    mapping of the vertices of $P_N$ to a path $P_M$, such that adjacent vertices are mapped to adjacent or equal vertices, and such that the total weight of the image of a vertex from $P_M$ equals an integer $c$. We show that {\sc Weighted Path Emulation}, with 
    $c$ as parameter, is hard for $W[t]$ for all $t\in {\bf N}$, and is strongly NP-complete.
    We also show that {\sc Directed Bandwidth} is hard for $W[t]$ for all $t\in {\bf N}$, for directed
    acyclic graphs whose underlying undirected graph is a caterpillar.
\end{abstract}

\section{Introduction}
The {\sc Bandwidth} problem is one of the classic problems from algorithmic graph theory. In this problem, we are given an undirected graph $G=(V,E)$ and integer $k$, and want to find a bijection from $V$ to
$\{1, 2, \ldots, n\}$, with $n=|V|$, such that for each edge $\{v,w\}\in E$: $|f(v)-f(w)| \leq k$.
The problem was proved to be NP-complete in 1976 by Papadimitriou~\cite{Papadimitriou76}. Later, several
special cases were proven to be NP-complete. In 1986, Monien~\cite{Monien86} showed that {\sc Bandwidth} stays NP-complete when the input is restricted to caterpillars with hair length at most three. A caterpillar is a tree where
all vertices of degree at least three are on the same path; the hairs are the paths attached to this central path, and have here at most three vertices.

In this paper, we consider the parameterized complexity of this problem. We consider the standard parameterization, i.e., we ask for the complexity of {\sc Bandwidth} as a function of $n$
and $k$. This problem is long known to belong to XP: already in 1980, Saxe~\cite{Saxe80} showed that {\sc Bandwidth} can be solved in time
$O(f(k)n^{k+1})$ for some function $f$; this was later improved to $O(f(k)n^k)$~\cite{GurariS84}.

In 1994, Bodlaender et al.~\cite{BodlaenderFH94} reported that {\sc Bandwidth} is $W[t]$-hard for all
positive integers $t$, even when we restrict the input to trees. However, the proof of this fact was so far never published. In the current paper, we give the proof of a somewhat stronger result: {\sc Bandwidth} is $W[t]$-hard for all 
positive integers $t$, even when we restrict the input to
caterpillars with maximum hair length three. 
A sketch of a proof that {\sc Bandwidth} is $W[t]$-hard for all
positive integers $t$ for general graphs appears in the monograph by Downey and Fellows \cite{DowneyF99}.
In recent years, Dregi and Lokshtanov~\cite{DregiL14} gave a proof that {\sc Bandwidth} is
$W[1]$-hard for trees of pathwidth at most two, and showed that there does not exist an
algorithm for {\sc Bandwidth} on such trees with running time 
of the form $f(k) n^{o(k)}$ assuming that the
Exponential Time Hypothesis holds. 

Our proof uses techniques from the 
NP-hardness proof for {\sc Bandwidth} on caterpillars by Monien~\cite{Monien86}. In particular, one gadget
in the proof is identical to a gadget from Moniens proof. Also, the proof is inspired by ideas
behind the proof of the result reported in \cite{BodlaenderFH94}, and a proof for $W[t]$-hardness of
a scheduling problem for chains of jobs with delays, which was obtained by Bodlaender and van der Wegen \cite{BodlaendervdW20}. 

To obtain our main result, we obtain an intermediate result that is also interesting on itself. We consider
a variation of the notion of uniform emulation. The notion of emulation was introduced by Fishburn and Finkel~\cite{FishburnF82}, to describe the simulation of processor networks on smaller processor networks.
An emulation of a graph $G=(V,E)$ on a graph $H=(W,F)$ is a function $f: V \rightarrow E$, such that
for each edge $\{v,w\}\in E$, $f(v)=f(w)$ or $\{f(v),f(w)\}\in F$, i.e., neighboring vertices are mapped to the same or neighboring vertices.
An emulation is uniform when each vertex in $H$ has the same number of vertices mapped to it, i.e.,
there is a constant $c$, called the {\em emulation factor}, such that for all $w\in W$: $|f^{-1}(w)|=c$.
An analysis of the complexity to decide whether for given $G$ and $H$, there exists a uniform emulation
was made by Bodlaender and van Leeuwen~\cite{BodlaenderL86}, and Bodlaender~\cite{Bodlaender90}. In particular, in \cite{Bodlaender90}, the complexity of deciding if there is a uniform emulation on a path or cycle was studied. It was shown that {\sc Uniform Emulation on a Path} belongs to XP, parameterized by the emulation factor $c$, belongs to XP for connected graphs and is NP-complete,
even for $c=4$, when we allow that $G$ is not connected.
Bodlaender et al.~\cite{BodlaenderFH94} claimed that {\sc Uniform Emulation on a Path} is hard
for $W[t]$ for all positive integers $t$. In this paper, we show a variation of this result, where the input
is a {\em weighted path}. We name the problem of finding an uniform emulation of a weighted path to a
path {\sc Weighted Path Emulation}.
It is straightforward to modify the algorithm from \cite{Bodlaender90} to weighted graphs. This shows
that {\sc Weighted Path Emulation} belongs to XP, with the emulation factor as parameter.

There is a sharp distinction between the complexity of the {\sc Bandwidth} problem for
caterpillars with hairs of length at most two, and caterpillars with hairs of length three (or larger).
Assmann et al.~\cite{AssmannPSZ81} give a chacterization of the bandwidth for caterpillars whose hair length is at most two, and show that one can compute a layout of optimal width in $O(n \log n)$ time. This
contrasts with the NP-hardness and fixed parameter intractability for caterpillars with hairs of length three, by Monien~\cite{Monien86} and this paper.

This paper is organized as follows. In Section~\ref{section:definitions}, we give a number of definitions.
In Section~\ref{section:emulation}, we show hardness for the {\sc Weighted Path Emulation} problem. Section~\ref{section:bandwidth} gives the main result: hardness for the bandwidth of caterpillars
with hairs of length at most three. In Section~\ref{section:directedbandwidth}, we discuss a variation
of the proof, to obtain that {\sc Directed Bandwidth} is hard for $W[t]$ for all positive integers $t\in {\bf N}$, for directed acyclic graphs whose underlying undirected graph is a caterpillar with hair length one.
Some final remarks are made in Section~\ref{section:conclusions}.

\section{Definitions}
\label{section:definitions}
All graphs in this paper are considered to be simple and undirected. We assume that the reader is familiar with standard notions from graph theory and fixed parameter complexity (see e.g.~\cite{DowneyF99,DowneyF13,FlumG06}).

$P_n$ denotes the path graph with $n$ vertices. We denote the vertices of $P_n$ by the first $n$ positive integers, $1, 2, \ldots, n$; the edges of $P_n$ are the pairs $\{i,i+1\}$ for $1 \leq i < n$.

A {\em caterpillar} is a tree such that there is a path that contains all vertices of degree at least three. A caterpillar can be formed by taking a path $P_N$ (the {\em spine}), and then attaching to vertices of $P_N$
zero or more paths. These latter paths are called the {\em hairs} of the caterpillar.

A {\em linear ordering} of a graph $G=(V,E)$ is a bijective function $f: V \rightarrow \{1, 2, \ldots, n\}$.
The {\em bandwidth} of a linear ordering $f$ of $G$ is $\max_{\{v,w\}\in E} |f(v)-f(w)|$. The {\em bandwidth} of a graph is the minimum bandwidth over its linear orderings.

Let $G=(V,E)$ be an undirected graph, and $w: V \rightarrow {\bf Z}^+$ be a function that assigns to each vertex a positive integer weight.
An {\em emulation } of $G$ on a path $P_M$ is a function $f: V\rightarrow \{1, 2, \ldots, M\}$, such that 
for all edges $\{v,w\}\in E$, $|f(v)-f(w)|\leq 1$. An emulation $f: V\rightarrow \{1, 2, \ldots, M\}$
is said to be {\em uniform}, if there is an integer $c$, such that
for all $i \in \{1,2, \ldots, r\}$, $\sum_{v: f(v)=i} w(v)=c$.
$c$ is called the {\em emulation factor}. 

For a directed acyclic graph $G=(V,A)$, the {\em directed bandwidth} of a topological ordering of $G$ is
$\max_{(v,w)\in A} f(w)-f(v)$; the {\em directed bandwidth} of a directed acyclic graph $G$ is the minimum directed bandwidth over all topological orderings of $G$.

If we have a directed graph $G=(V,A)$, the {\em underlying undirected graph} of $G$ is the
undirected graph $G'=(V,E)$, with $E= \{ \{v,w\} ~|~ (v,w)\in A\}$; i.e., we forget the direction of edges.

We consider the following parameterized problems. 

\begin{verse}
{\sc Bandwidth}\\
{\bf Given:} An undirected graph $G=(V,E)$, integer $k$.\\
{\bf Parameter:} $k$.\\
{\bf Question:} Is the bandwidth of $G$ at most $k$?
\end{verse}

\begin{verse}
{\sc Directed Bandwidth}\\
{\bf Given:} A directed acyclic graph $G=(V,E)$, integer $k$.\\
{\bf Parameter:} $k$.\\
{\bf Question:} Is the directed bandwidth of $G$ at most $k$?
\end{verse}

\begin{verse}
{\sc Weighted Path Emulation}\\
{\bf Given:} Integers $N$, $M$, $c$, weight function $w: \{1, 2, \ldots, N\} \rightarrow {\bf Z}^+$,  such that $\sum_{i=1}^N w(i) / M = c \in {\bf Z}^+$.\\
{\bf Parameter:} $c$.\\
{\bf Question:} Is there a uniform emulation $f$ of $P_N$ with weight function $w$ on $P_M$.
\end{verse}
Note that in the problem statement above, $c$ is the emulation factor, i.e., we have for a solution $f$
that for each
$j$, $1\leq j\leq M$, $\sum_{i:f(i)=j} w(i) = c$.

A Boolean formula is said to be {\em $t$-normalized}, if it is the conjunction of the disjunction of the conjunction of \ldots of literals, with $t$ alternations of AND's and OR's. So, a Boolean in Conjunctive 
Normal Form is 2-normalized. Downey and Fellows~\cite{DowneyF95} consider the following parameterized problem; this is
the starting point for our reductions.

\begin{verse}
{\sc Weighted $t$-Normalized Satisfiability}\\
{\bf Given:} A $t$-normalized Boolean formula $F$ and a positive integer $k\in {\bf Z}^+$. \\
{\bf Parameter:} $k$\\
{\bf Question:} Can $F$ be satisfied by setting exactly $k$ variables to true?
\end{verse}

\begin{theorem}[Downey and Fellows \cite{DowneyF95}]
{\sc Weighted $t$-Normalized Satisfiability} is $W[t]$-complete.
\end{theorem}

\section{Hardness of {\sc Weighted Path Emulation}}
\label{section:emulation}

In this section, we show that the {\sc Weighted Path Emulation} problem is $W[t]$-complete
for all positive integers $t\in {\bf N}$. Some techniques of this proof are inspired from
techniques from a similar proof for a scheduling problem by Bodlaender and van der Wegen~\cite{BodlaendervdW20} and the proof underlying some of the results reported in \cite{BodlaenderFH94}, see also \cite{DowneyF99}.

Suppose we are given a $t$-normalized Boolean expression $F$ over $n$
variables, say $x_1, \ldots, x_n$, and integer $k$. We let $t'$ be the number of nested levels
of disjunction. We consider the problem to satisfy $F$ by making exactly $k$ variables true.

We will define a path $P_N$ with a weight function $w: \{1, \ldots, N\} \rightarrow {\bf Z}^+$, an emulation factor $c$,
and an integer $M$, such that $P_N$ has a uniform emulation on a path $P_M$ if and only if $F$ can be satisfied by setting exactly $k$ variables to true. Before giving the proof, we give a high level overview of some main ideas of the proof.

\subsection{Intuition and techniques}

In this subsection, we give some ideas behind the construction. The precise construction and formal proofs are given in the next subsection.

We assume we have given a $t$-normalized Boolean formula $F$. We transform the formula to a weighted path $P_M$,
such that $P_M$ has a uniform emulation on $P_M$ with $c$ the emulation factor, if and only if $F$
can be satisfied by setting exactly $k$ variables to true.

We can view $F$ as a tree, with internal nodes marked with disjunction or conjunction, and each leaf
with a literal, then we alternatingly have a level in the tree with disjunctions, and with conjunctions.
We set $t'$ to be the number of levels with disjunctions.

The path $P_M$ is formed by taking, in this order, the following: a part called the
`floor', $k$ `variable parts', $t'$ `disjunction parts', and 
a `filler path'. $t'$ is the number of levels in the formula tree with disjunctions, and for each `level'
of disjunction we have one disjunction part. E.g., if $F$ is in conjunctive normal form, then $t'=1$.

The {\em floor} has $M$ vertices, each with a weight that is larger than $c/2$. Thus, we cannot map two
floor vertices to the same element of $P_M$, and thus, can assume, without loss of generality, that
the $i$th floor vertex is mapped to $i$. 
The different weights for floor vertices help to build the further gadgetry of the construction.

\begin{figure}[htb]
    \centering
    \includegraphics{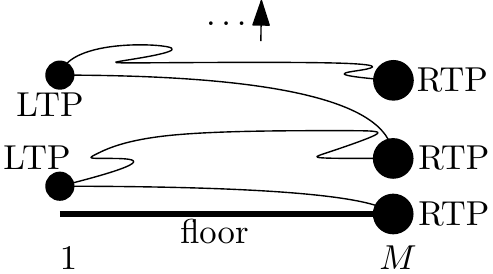}
    \caption{Impression of a first part of the construction. The $i$th floor vertex is mapped to $i$. The path then moves from $M$ to $1$ and back, with left turning points (LTR) mapped to 1, and right turning points (RTP) mapped to $M$. The picture shows the floor and first two variable parts.}
    \label{figure:intuition-turning}
\end{figure}

The variable and disjunction parts are forced to start at $M$, then move to $1$, and then move (possibly with some `zigzagging') back to $M$, where then the next part starts. This is done by giving each
part one vertex of large weight that only can fit at vertex $1$, and another vertex of even larger weight, that only can fit at vertex $M$. These large weight vertices are called left and right turning points.
See Figure~\ref{figure:intuition-turning} for an impression of the construction.

We have $k$ {\em variable parts}. Each models one variable that is set to true. We start with
a left turning point, $M-2$ vertices of weight one, and a right turning point: this is to move back
from $M$ to $1$. Then, we have $n-1$ vertices of weight one, $M-2n-2$ `heavy' vertices, and again
$n-1$ vertices of weight one. The heavy vertices are mapped consecutively (except possibly at the first
$n$ and last $n$ positions); the weight one vertices before and after the heavy vertices allows us
to shift the sequence of heavy vertices in $n$ ways --- each different such shift sets another variable to
true. By using two different heavy weights, combined with weight settings for floor vertices and
vertices from disjunction parts, we can check that all variable parts select a different variable
to be true (which is done at positions $n+2, \ldots, 2n+1$), and that $F$ is satisfied (which is
done at positions $2n+2, \ldots, M-n-2$).

We have for each level in the formula tree with disjunctions a {\em disjunction part}. Thus, we have
$t'$ disjunction parts.
With help of `anchors' (vertices of large weight that can go only to one specific position), we
can ensure that a subpart for a disjunction has to be mapped to the part of the floor that corresponds with this disjunction. Such a subpart consist of a path with weight one vertices, a path with
$3m(F')$ {\em selecting vertices} (which have larger weight), and another path with weight one vertices.
Now, each term in the disjunction has an associated interval of size $m(F')$ and between these intervals
we have $m(F')$ elements. Then, we can show that the selecting vertices must cover entirely one
of the intervals of a term --- this corresponds to that term being satisfied. See Figure~\ref{figure:3cover1}, and the proof of Claim~\ref{claim:selecteddisjunctionterm} for an
illustration and details.

Heavy vertices of variable parts come in two weights: $c^v$ and $c^c+c^u$.
This is
used for checking that $F$ is satisfied. As an example, consider a negative literal $\neg x_j$ in $F$.
We have one specific position on $P_M$, say $i$,
that checks whether this literal is satisfied, in case its
satisfaction contributes to the satisfaction of $F$ --- that case corresponds to having a selecting vertex mapped to $i$ for each
level of disjunction. Now, the weight of the floor vertex mapped to $i$
is such that when this floor vertex and all selecting vertices are mapped to $i$, then we can only
fit $k$ heavy vertices of weight $c^v$ here; if at least one of these heavy vertices has weight $c^v+c^u$,
then the total weight mapped to $i$ exceeds $c$. If this happens, then this heavy vertex belongs to
a variable part which corresponds to setting 
$x_i$ to be true; thus, this enforces that $x_i$ is false. A somewhat similar construction
is used for positive literals.

\begin{figure}
    \centering
    \includegraphics{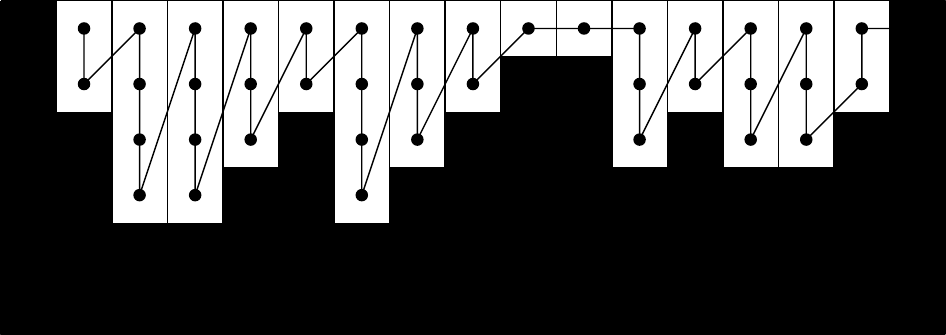}
    \caption{Illustration of the mapping of the filler path. The black area represents the weights of
    floor, variable part and disjunction part vertices mapped to the element of $P_M$}
    \label{figure:fillepath}
\end{figure}

The last part of $P_N$ is the {\em filler path}. This is a long path with vertices of weight one.
This is used to ensure that the mapping becomes uniform: if the total weight of vertices of floor,
variable part, and disjunction parts vertices mapped to $i$ is $z_i$, then we map $c-z_i$ (consecutive) vertices of the filler path to $i$. See Figure~\ref{figure:fillepath} for an illustration.

\subsection{Proof of hardness of {\sc Weighted Path Emulation}}

\paragraph{Elements, weights, and intervals}
A $t$-normalized Boolean formula $F$ can be represented by an ordered rooted tree with $t+1$ levels. Each node in the tree represents a subformula of $F$. Each leaf represents a literal, which can be positive or
negative. Each internal node represents the conjunction or disjunction of the formulas represented by the
children of the nodes. The formula represented by the root is $F$. We call the formulas represented
by each of the nodes of this tree the {\em elements} of $F$; e.g., each occurrence of a literal is an element, $F$ itself is an element, etc. 
Note that the elements in $F$ are nested. We also use terminology of rooted trees for the elements, like children, ancestors, descendants, in the natural sense.

We associate to each element an {\em interval size}, which we denote
by $s(F')$ for the interval size of element $F'$.
If $F'$ is a literal, then $s(F') = 2n+1$. 

The interval size of a conjunction $F' = F_1 \wedge F_2 \wedge \cdots \wedge F_q$
is the sum of the interval sizes of its terms, i.e., $s(F')= s(F_1 \wedge F_2 \wedge \cdots \wedge F_q) = \sum_{i=1}^q s(F_i)$. 

For a disjunction $F' = F_1 \vee F_2 \vee \cdots \vee F_q$, we define 
$m(F')= \max_{1\leq i\leq q} s(F_i)$.
(As $F$ is normalized, we have actually that $s(F_1) = s(F_2) = \cdots = s(F_q)$.) So, $m(F')$ gives the size of the elements of which $F'$ is a conjunction.
Now, we set $s(F') = (10q+5)\cdot m(F)$. 

We will assign to each 
element $F'$ an interval $[\ell(F), r(F)]$. We define these recursively, top-down.
We have that each interval of an element $F'$ contains exactly $s(F')$ integers 
(i.e., $r(F')-\ell(F')+1 = s(F')$), and the nesting and
intersections of intervals of elements follow precisely how these are nested in the formula $F$.
In addition, for disjunctions, we define {\em left anchor points} and {\em right anchor points}. These
anchor points are inside the interval assigned to the disjunction --- note that the anchor points
are different from the left and right side of the interval.

For $F$, we set $\ell(F)= 2n+2$, and $r(F)= 2n+1 + s(F)$.

Suppose $F'$ is a conjunction $F'= F_1 \wedge F_2 \cdots \wedge F_q$. 
Assign to $F_1$ the interval $[\ell(F), \ell(F)+s(F_1)-1]$, to $F_2$ the interval $[\ell(F)+s(F_1),
\ell(F)+s(F_1)+s(F_2)-1]$, etc. I.e., $\ell(F_i) = \ell(F) +\sum_{j=1}^{i-1} s(F_j)$ and
$r(F_i) = \ell(F_i)+s(F_i)-1$. 

Suppose $F'$ is a disjunction $F'= F_1 \vee F_2 \cdots \vee F_q$. 
Set the {\em left anchor point} of $F'$ to $LA(F')=\ell(F')+ (4q+2)\cdot m(F')$, and the {\em right anchor point} of $F'$ to
$RA(F')=\ell(F')+(6q+3)\cdot m(F')$. The intervals of the terms $F_i$ will be between these anchor points:
we assign to $F_i$ the interval $[\ell(F'),r(F')]=[LA(F')+ (2i-1) \cdot m(F'), LA(F')+ 2i \cdot m(F') -1]$.

\begin{figure}[htb]
    \centering
    \includegraphics[width=\textwidth]{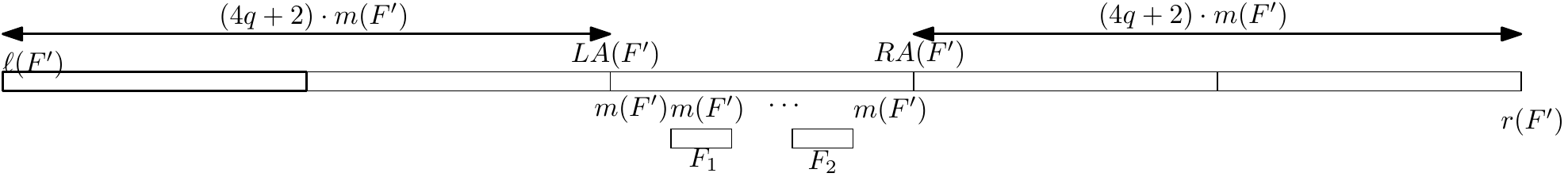}
    \caption{An example of the nested intervals of disjunctions}
    \label{figure:nestedintervals}
\end{figure}

One can note that each element $F'$ has an interval containing $s(F')$ integers. 
For a literal $F'$, we say that $\ell(F')+n$ is the {\em midpoint} of its interval; indeed, the midpoint is the middle element of the interval of the literal, as $s(F')=2n+1$.

Also note that the intervals of elements are nested, with the same nesting as these elements have in the formula. See Figure~\ref{figure:nestedintervals} for an illustration.

We also associate to each element an integer, called its {\em or-depth}. 
If $F$ is a conjunction, its or-depth is $0$; if $F$ is a disjunction, its or-depth is $1$.
The or-depth of a literal or conjunction equals the or-depth of its parent in the formula tree;
the or-depth of a disjunction is exactly one larger than the or-depth of its parent in the formula tree.
Note that the maximal or-depth of an element is at most $\lceil t/2 \rceil$, and because the formula
is normalized, all terms of a conjunction and of a disjunction have the same or-depth.
All conjunctions with the same or-depth form a {\em level} of the formula tree, as do all
disjunctions with the same or-depth.

We set $M = 2+ 3n+s(F)$.

\paragraph{Definition of constants}
We now define a series of constants. Note that each of these only depends on $k$ and $t$. These constants are used in the definition of the weights of vertices.

We have a number of constants:
\begin{itemize}
    \item $c^p=5t'+3k+1$. This denotes the space for vertices of width 1; we have several parts of the path with such vertices. (The $p$ denotes path.)
    \item $c^u = 3 \cdot c^p$. 
    \item We have $t'$ constants that denote weights for vertices in parts that check that a disjunction
    holds. These weights are so large that where this is necessary, no two vertices of the same part
    fit together, and thus that of such a part, when these vertices are consecutive, they will be mapped
    to consecutive vertices.
    We set: $c^d_{t'} = 3\cdot k \cdot c^u$, and for $1\leq i < t'$: $c^d_i = 3 \cdot c^d_{i+1}$.
    \item A similar role, for the choice of variables that are true, plays the constant $c^v$.
    We set $c^v = 3 \cdot c^d_{t'}$.
    \item We have $t$ constants that denote weights that ensure that some vertices of disjunction paths
    are mapped to specific vertices on $P_M$. These `anchor' these vertices; we have
    variables $c^a_1, \ldots, c^a_{t'}$ with $c^a_{t'} = 3 \cdot c^v$ and for $1 \leq i < t'$: $c^a_i = 3 \cdot c^a_{i+1}$.
    \item We have very large constants for the turning points. For the left turning point,
    we have $c^L = 2c^a_{t'}$ and for the right turning point $c^R = (k+t'+1) \cdot c^L$.
    \item Finally, we set $c = 2 \cdot (k+t') \cdot c^R+1$.
\end{itemize}

Note that:
\[
1 < c^p < c^u < c^d_{t'} < 
\cdots < c^d_2 < c^d_{1} < c^v < c^a_{t'} < \cdots < c^a_2 <  c^a_{1} < c^L < c^R < c
\]


Recall that $M = 2+ 3n+s(F)$. The value of $N$ is not given explicitly, but follows from the 
construction. The emulation factor equals $c$.
From the definitions above, we can see
that $c$ only depends on $k$ and $t'$, and thus is a function of the parameter.

\paragraph{The weight function $w$}
We now define the weight function $w: \{1, \ldots, N\} \rightarrow {\bf Z}^+$.

We name the successive different parts of $P_N$ as follows: {\em floor}, {\em true variable parts},
{\em disjunction  parts}, and {\em filler path}.

\subparagraph{Floor}
The floor consists of the first $M$
vertices of $P_N$. These vertices have the following weights.
\begin{itemize}
    \item We have $w(1)= c- (k+t') c^L$. (Left turning point.)
    \item For $2\leq i\leq n+1$, $w(i)= c - k \cdot c^v - c^p$. (Left part of variable selection.)
    \item For $n+2 \leq i \leq 2n+1$, $w(i) = c - k \cdot c^v - c^u - c^p$. (Part that verifies that no variable is selected twice.)
    \item 
          For $2n+2 \leq i \leq M-n-1$, we have a simple procedure that computes $w(i)$:
          \begin{itemize}
              \item Start by setting $w(i)= c - c^p - k \cdot c^v$.
              \item If $i$ is a left or right anchor point of a disjunction $F$, then suppose this disjunction has or-depth $\alpha$. Now, subtract $c^a_\alpha$ from the current value of $w(i)$.
              \item If there is a disjunction of or-depth $\alpha$, such that $i$ is between the left and right anchor point of this disjunction, then 
              subtract $c^d_\alpha$ from the current value of $w(i)$. This step is executed for each applicable $\alpha$.
              \item If $i$ is the midpoint of the interval of an element that is a positive literal then subtract $(k-1) \cdot c^u$ from the current value of $w(i)$. 
              \item If $i$ is not the midpoint of any interval of an element that is a literal, then
              subtract $k \cdot c^u$ from the current value of $w(i)$.
          \end{itemize}
          This part is for verification that the variable selection satisfies $F$.
    \item For $M-n \leq i \leq M-1$,  $w(i)= c - k \cdot c^v - c^p$. (Right part of variable selection.)
    \item We have $w(M)= c- (k+t') c^R$. (Right turning point.)
\end{itemize}

The intuition behind the construction
of the floor is the following. The $i$th vertex of the floor is mapped to the $i$th vertex
of $P_M$. Thus, the other vertices mapped to $i$ must have a total weight of $c-w(i)$.

\subparagraph{True variable parts}
After the floor, we have $k$ parts, called {\em true variable parts}. The first vertex of the first
true variable part is adjacent to the last vertex of the floor; the last vertex of the $i$th true variable
part is adjacent to the first vertex of the $(i+1)$st true variable part ($1\leq i < k$).

For $i$ from $1$ to $k$, we take the following vertices with weights, successively, forming the
$i$th true variable part.
\begin{itemize}
    \item A vertex of weight $c^R$. (This vertex will be mapped to $M$; the {\em right turning point}.)
    \item $M-2$ vertices of weight $1$. (These vertices go back from $M$ to $1$.)
    \item A vertex of weight $c^L$. (This vertex will be mapped to $1$; the {\em left turning point}.)
    \item $n-1$ vertices of weight $1$. (These vertices are mapped with possibly a fold from 1 to a vertex in $\{2, \ldots, n\}$. This ensures that the first vertex of the next subpart is mapped to a vertex
    in $\{2, \ldots, n+1\}$.)
    \item $2n+s(F)$ vertices of weight either $c^v$ or $c^v + c^u$. We call these the {\em heavy} vertices of the true variable part. The following vertices have
    weight $c^v + c^u$; all others have weight $c^v$:
    \begin{itemize}
        \item The $n$th vertex of this subpart, i.e., we first have $n-1$ vertices of weight $c^v$ and
        then one of weight $c^v+c^u$. (This latter vertex is mapped to a vertex in
        $\{n+2, 2n+1\}$, the part to check that no variable is selected once. Mapping to $n+1+j$ corresponds to setting $x_j$ to true.) Call this the {\em determining} vertex of the part.
        \item For each element $F'$ of $F$ that is a positive literal $x_i$, for each integer
        $j$ in the interval of $F'$, (i.e., 
        in $[\ell(F'),r(F')-1]$) that is not the midpoint of this interval,
        (i.e., $j\neq \ell(F)+n$), we have that the
        $(j-i-1)$th vertex of this subpart has weight $c^v + c^u$.
        \item For each element $F'$ of $F$ that is a negative literal $x_i$, for the midpoint
        $j$ of the interval of $F'$, we have that the 
        $(j-i-1)$th vertex of this subpart has weight $c^v + c^u$.
    \end{itemize}
    The role of these vertices is in checking whether literals are satisfied.)
    \item $n-1$ vertices of weight 1. (The right turning point of the next subpart is mapped to $M$; these vertices make the connection of the last vertex defined above to this right turning point of the next subpart.)
\end{itemize}

\paragraph{Disjunction parts}
We have $t'$ disjunction parts, each for of the $t'$ different or-depths of disjunctions. For $\alpha=1$ to $t'$, we have
a part that handles all disjunctions of or-depth $\alpha$. 
For $\alpha$ from $1$ to $t'$ we have the following successive vertices with weights:
\begin{itemize}
    \item A vertex of weight $c^R$. (Again, this right turning point goes to $M$.)
    \item $M-2$ vertices of weight 1. (These vertices go back from $M$ to $1$.)
    \item A vertex of weight $c^L$. (Again, this is the left turning point and goes to $1$.)
    \item Consider the elements of $F$ that are a disjunction of or-depth $\alpha$. Name these $F_1, \ldots, F_q$
    and assume they appear in $F$ in this order, from left to right. Note that 
    $LA(F_1) < RA(F_1) < LA(F_2) < RA(F_2) < \cdots < LA(F_q) < RA(F_q)$.  Each $F_i$ is a disjunction
    of elements that all have the same size $m(F_i)$.
    To avoid a case analysis for the first element, write $RA(F^0)=1$. 
    For $j$ from $1$ to $q$, we have successively:
    \begin{itemize}
        \item $LA(F_j)-RA(F_{j-1})-1$ vertices of weight $1$. (These go from the right anchor point of the previous disjunction (or from 1) to the left anchor point of this disjunction.) 
        \item A vertex of weight $c^a_\alpha$. This 
        vertex is called the {\em left anchor} of the disjunction. We will show later that a left anchor must be mapped to the corresponding left anchor point.
        \item $(2q-2)\cdot m(F_i)$ vertices of weight 1. (This goes from the left anchor point to the next subpart. It is possibly folded.) 
        \item $3m(F_i)$ vertices of weight $c^d_{\alpha}$. This subpart is to select an element of the disjunction that is satisfied. Call these vertices the {\em selecting} vertices of the disjunction.
        \item $(2q-2)\cdot m(F_i)$ vertices of weight 1. (This goes from the previous subpart to the right anchor point. It is possibly folded.) 
        \item A vertex of weight $c^a_\alpha$. This 
        vertex is called the {\em right anchor} of the disjunction. We will show later that a right anchor must be mapped to the corresponding right anchor point.
    \end{itemize}
    \item $M-RA(F_q)-1$ vertices of weight 1. (These go from the right anchor point of $F^q$ to $M$.)
\end{itemize}

\paragraph{Filler path}
Count the total weight of all vertices, defined so far. Suppose this is $\gamma$. Take $cM-\gamma$ vertices
of weight $1$. This is the last part of $P_N$, and finishes the construction of $P_N$; we do not give an explicit formula for $N$,
but just state that $N$ is the number of vertices of all parts of this path as defined above.
With help of the filler path, the total weight of all vertices is exactly $cM$.

(The first vertex of the filler path will be mapped to $M-1$. The construction is such that $1$ and $M$
will have $c$  weight mapped to it, by floor and turning point vertices;
all other vertices (between $1$ and $M$) have
less than $c$ weight mapped to it by all vertices except filler path vertices. The filler path thus is used to `fill the remaining gaps' after all other
vertices are mapped, ensuring that each vertex on $P_M$ has exactly weight $c$ mapped to it.)

\paragraph{Correctness proof}

\begin{lemma}
If there is a uniform emulation of $P_N$ to $P_M$ with weight function $w$, then $F$ can be satisfied by setting exactly $k$ variables to true.
\label{lemma:snake2formula}
\end{lemma}

\begin{proof}
Let $f$ be a uniform emulation of $P_N$ to $P_M$. We show the lemma with help of a number of claims.

\begin{claim}
One of the following two must be the case:
\begin{itemize}
    \item For each $i$, $1\leq i\leq M$, the $i$th floor vertex is mapped to $i$, i.e., $f(i)=i$.
    \item For each $i$, $1\leq i\leq M$, the $i$th floor vertex is mapped to $M-i+1$, i.e., $f(i)=M-i+1$.
\end{itemize}
\end{claim}

\begin{proof}
Note that for each floor vertex (vertices $1, 2, \ldots , M$ of $P_N$) has a weight that is larger than
$c/2$. So, we cannot map two floor vertices to the same vertex from $P_M$. As there are $M$ floor
vertices, each vertex in $P_M$ has one floor vertex mapped to it.
\end{proof}

Without loss of generality, we suppose that the first is the case, i.e., for each $i$, $1\leq i\leq M$, $f(i)=i$. (The second case can be transformed to the first by using the 
the function $f'$, with for all $v$, $f'(v)= M+1-f(v)$.)


\begin{claim}
All right turning point vertices are mapped to $M$. All left turning point vertices are mapped to $1$.
\end{claim}

\begin{proof}
For all $i\neq M$, the $i$th floor vertex
has a weight that is larger than $c-c^R$, and is mapped to $i$. Thus, we cannot map a right turning point (whose weight is $c^R$) to such a vertex. So, necessarily, right turning points are mapped to $M$.

For all $i\not\in\{1,M\}$, the $i$th floor vertex has a weight that is larger than $c-c^L$ and is mapped
to $i$. Hence, we cannot map left turning points (who have weight $c^L$) an integer $i\not\in \{1,M\}$.
We also cannot map it to $M$, as the total weight of the floor vertex mapped to $M$ and all right turning points equals $c$.
So, all left turning points must be mapped to $1$.
\end{proof}

\begin{claim}
For all elements $F'$ of $F$ that are a disjunction, the left anchor point of $F'$ is mapped to
the left anchor of $F'$, $LA(F')$ and the right anchor point of $F'$ is mapped to $RA(F')$.
\label{claim:disjunctionanchors}
\end{claim}

\begin{proof}
We prove this by induction to the or-depth of the disjunction. Suppose the claim
holds for all disjunctions of or-depths smaller than $\alpha$. 


The anchors of disjunctions at or-depth $\alpha$ have weight $w^a_\alpha$. Thus, if such an anchor
is mapped to $i$, the total weight of the floor vertex, turning points, and anchors of disjunctions
of or-height smaller than $\alpha$ must be at most $c-w^a_\alpha$. One can observe that the
only elements that fulfill this are the 
are the anchor points of or-depth $\alpha$, i.e., vertices on $P_M$ of the type $LA(F'')$ or $RA(F'')$ for disjunctions
$F''$ of or-depth $\alpha$. None of these has at least $2 \cdot w^a_\alpha$ weight left, and
the number of such vertices on $P_M$ equals the number of anchor points on $P_N$. So, these anchors
must be mapped one to one to these anchor points, of the same or-depth.

Now, with induction from left to right, we see that the left and right anchors of a disjunction
at or-depth $\alpha$ are mapped to the corresponding anchors. 
Say the disjunctions of or-depth $\alpha$ are $F_1$, $F_2$, \ldots, $F_q$, from left to right.
We now prove the result, with induction from left to right. Assume that the left and right
anchors of $F_1$, \ldots, $F_{j-1}$ are mapped to their corresponding anchor points, for some $j$.
Consider $F_j$. The distance on $P_N$ from the left anchor of $F_j$ to the right anchor of $F_{j-1}$
(or the left turning point of the $\alpha$th level of disjunctions, in case $j=1$) equals the
distance on $P_M$ of the left anchor point of $F_j$ to the right anchor point of $F_{j-1}$ (or to $1$, in
case $j=1$); on $P_M$ there are no other anchor points of this or-depth level between these. So,
the left anchor of $F_j$ must be mapped to $LA(F_j)$.
The distance on $P_N$ from the left anchor of $F_j$ to the right anchor of $F_j$ equals
$(4q-1)\cdot m(F_j)$. The right anchor of $F_j$ can fit in the right anchor point of $F_j$, but it
cannot go to another anchor point: in $(4q-1)\cdot m(F_j)$ steps, one cannot move from the left anchor
point to any element outside the interval of $F_j$ (see Figure~\ref{figure:nestedintervals}.)
\end{proof}

\begin{claim}
Let $n+1 \leq j \leq M - n$, and let $1\leq i\leq k$. The $i$th variable part has at least one heavy vertex mapped to $j$.
\label{claim:oneheavy}
\end{claim}

\begin{proof}
The left turning point of the variable part is mapped to $1$; the last vertex of the part is mapped to
a right turning point which is mapped to $M$. Thus, the last of the $n-1$ vertices of weight 1 that come after the left turning point in this part
is mapped to an integer that is at most $n$, and the first of the last $n-1$ vertices of weight 1 in the part is mapped to an integer that is at least $M - n+1$. Between these two vertices, the variable part has a path with only heavy vertices; thus, all integers that are larger than
$n$ and smaller than $M-n+1$ have at least one of these heavy vertices mapped to it.
\end{proof}

\begin{claim}
Let $n+1 \leq j \leq M - n$, and let $1\leq i\leq k$. The $i$th variable part has exactly one heavy vertex
mapped to $j$.
\end{claim}

\begin{proof}
First, suppose that $2n+2 \leq j \leq M-n-1$. 
Consider the total of the floor vertex mapped to $j$, vertices from disjunction  parts mapped to $j$, and one heavy vertex per true variable part (see Claim~\ref{claim:oneheavy}). 
Using earlier claims, we see that this total weight is at least 
$c - \sum_{\alpha=1}^{t'} c^d_\alpha + k \cdot c^u + c^p > c - c^v$. 
Thus, we cannot map a second heavy vertex of a variable part to $j$.

The cases that $n+1 \leq  j \leq 2n+1$ and $j=M-n-2$ can be shown with a similar counting argument.
\end{proof}

We now define a truth assignment to variables $x_1, \ldots, x_n$, as follows. We set $x_i$ to true,
if there exists a true variable part whose determining vertex is mapped to $n+1+i$, and to false otherwise.

\begin{claim}
Exactly $k$ variables are set to true.
\label{claim:exactktrue}
\end{claim}

\begin{proof}
As we have $k$ true variable parts, at most $k$ variables are set to true. 
If two true variable parts select the same variable, then the determining vertices of these parts have
the same image. The weight left at this image was $c^u+c^p$, and as $c^u<c^p$, we cannot fit twice
the extra weight of determining vertices here. So, each true variable part selects a different variable to be true.
\end{proof}

We now give a top-down recursive definition of which elements of $F$ are {\em selected}:
\begin{itemize}
    \item $F$ is selected.
    \item If $F'$ is a selected conjunction, then all its terms are selected.
    \item If $F'$ is a selected disjunction, then a term $F''$ of $F'$ is selected, when each integer
    in the interval of $F''$ (i.e., $[\ell(F''),r(F'')]$) has a selecting vertex of $F'$ mapped to it.
\end{itemize}

\begin{claim}
Suppose element $F'$ is selected, and has or-depth $\alpha \leq t'$. For each $\beta < \alpha$, each
integer in the interval of $F'$ has a selecting vertex of weight $c^d_{\beta}$ mapped to it.
\label{claim:ancestorselecting}
\end{claim}

\begin{proof}
From the definition of selection, it follows that each ancestor of $F'$ in the formula tree is selected.
If an ancestor of $F'$ has a disjunction as parent, then by definition, each integer in the interval
of that ancestor has a selecting vertex of the disjunction  part for that parent mapped to it; the
same holds for $F'$ itself. From this, the claim follows.
\end{proof} 

\begin{claim}
Each selected disjunction has at least one selected term.
\label{claim:selecteddisjunctionterm}
\end{claim}

\begin{proof}
Suppose that $F' = F'_1 \vee F'_2 \vee \cdots \vee F'_q$ is a selected disjunction of or-depth $\alpha$.
We consider the disjunction selection part of $F'$.

Recall from Claim~\ref{claim:disjunctionanchors} that the left anchor of $F'$ is mapped
to the left anchor point of $F'$, and the right anchor of $F'$ is mapped to the right anchor point.
The distance between these anchor points is $(2q+1)\cdot m(F')$; each of the terms $F'_i$ has interval
size $m(F')$. 

The first selecting vertex of the disjunction selection part of $F'$ has a path of length
$(2q+1)\cdot m(F')$ to the right anchor, namely with $3m(F')-1$ edges to the other selecting vertices 
of this part, 
$(2q-2)\cdot m(F')$ edges vertices of weight one, and one edge to the right anchor.
Hence, it cannot be mapped to an integer that is smaller than the left anchor point of the interval of
$F'$; all other selecting vertices of this part have a shorter path to the right anchor and hence
also cannot be mapped to an integer that is smaller than the left anchor point of the interval of $F'$.

With a similar argument, we can show that the selecting vertices of the disjunction selection part
cannot be mapped to an integer that is larger than the right anchor point of the interval of $F'$.
So, all selecting vertices are mapped to integers in $[LA(F'),RA(F')]$.

We cannot have two of these selecting vertices mapped to the same integer in $[LA(F'),RA(F')]$. 
This can be seen by considering the weight of all vertices mapped to one such integer; these include
the floor vertex, when the integer is an anchor point, the corresponding anchors, $k$ heavy vertices of true variable selecting parts, and (by Claim~\ref{claim:ancestorselecting}), for each $\beta$ smaller than or equal to the or-depth of $F'$,
a vertex of weight $w^d_\beta$. This gives a total weight of at least $c-c^p - (k-1)\cdot c^u - \sum_{\beta=1}^{\alpha} c^d_\beta$. By noting that $c^d_\alpha > c^p + (k-1)\cdot c^u + \sum_{\beta=1}^{\alpha} c^d_\beta$, we see that we cannot map two selecting vertices of disjunction
 parts of or-depth $\alpha$ to the same integer.
 
It thus follows, that the selecting vertices must be mapped to $3m(F')$ consecutive 
integers from $[LA(F'),RA(F')]$. This implies it must cover an interval $[LA(F'_j),RA(F'_j)]$ for some
$j$ entirely: each of these intervals has size $m(F')$ and between each pair of these,
we have $m(F')$ integers not in such intervals. See Figure~\ref{figure:3cover1}: if we map
the orange part to the interval between left and right anchor points, then it covers one
of the intervals of an $F_i$ (green in the figure) entirely.

\begin{figure}[htb]
    \centering
    \includegraphics{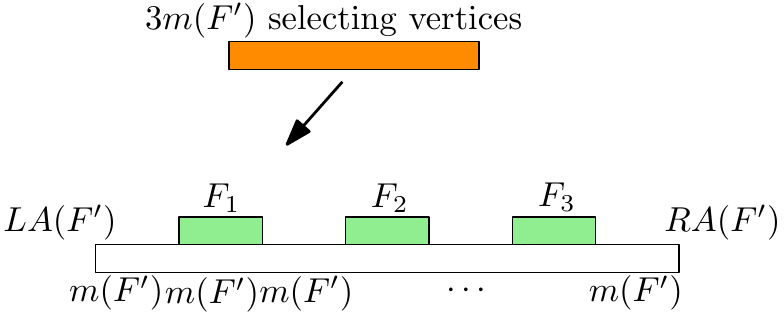}
    \caption{Illustration to the proof of Claim~\ref{claim:selecteddisjunctionterm}}
    \label{figure:3cover1}
\end{figure}

This implies that this term $F'_j$ is selected.
\end{proof}

\begin{claim}
Let $F'$ be a literal, and suppose the $j$th variable part selects $x_i$. There is one heavy vertex
of the $j$th variable part mapped to the midpoint of the interval of $F'$. The weight of this heavy vertex
equals
\begin{itemize}
    \item $c^d$, when $F'$ is the positive literal $x_i$, or when $F'$ is a negative literal
    $\neg x_{i'}$ with $i'\neq i$;
    \item $c^d+c^u$, when $F'$ is the negative literal $\neg x_i$, or when $F'$ is a positive literal
    $x_{i'}$ with $i'\neq i$.
\end{itemize}
\label{claim:literalblobs}
\end{claim}

\begin{proof}
Let $m$ be the midpoint of the interval of $F'$.
If the $j$th variable part selects $x_i$, then the determining vertex ($n$th heavy vertex)
is mapped to $n+i+1$, thus, the $r$th heavy vertex is mapped to $r+i+1$. Hence, to
$m$, the $m-i-1$st heavy vertex is mapped. The weight of this heavy vertex fulfills the properties
as stated in the claim, by construction.
\end{proof}

\begin{claim}
If $F'$ is a selected literal, then $F'$ holds under the defined truth assignment.
\label{claim:demonstratedliteral}
\end{claim}

\begin{proof}
When a literal is selected, then all its ancestors in the formula tree are selected. Thus,
each integer in the interval of the selected literal (and in particular, its midpoint) has
a heavy vertex of a disjunction  part of or-depth $\beta$ mapped to it, for
all $\beta \in [1, \ldots, t']$. The weight of all heavy vertices of disjunction  parts mapped to the midpoint of the interval of the literal equals $\sum_{\beta=1}^{t'} c^d_\beta$.

We first consider the case that $F'$ is a positive literal.
We look at the weights of all vertices mapped to the midpoint of the interval of $F$.
The floor vertex mapped to this midpoint has
weight $c - c^p - k\cdot c^v - \sum_{\beta=1}^{t'} c^d_\beta - (k-1)\cdot c^u$.

Now, suppose $F'$ does not hold. Then no true variable part
selects $x_i$, and thus, by Claim~\ref{claim:literalblobs},
each true variable part maps a vertex with weight of the form $c^v + c^u$ to the midpoint of
$F'$. Add these weights to the weight of the floor vertex, and heavy vertices of disjunction
 parts gives a total weight of at least $c - c^p + c^u > c$ mapped to the midpoint of the interval
of $F'$; contradiction. So, $F'$ holds.

Now, consider the case that $F'$ is a negative literal $\neg x_i$. The analysis is similar to the 
previous case. Now, the weight of the floor vertex mapped to the midpoint of the interval of $F'$
is $c - c^p - k\cdot c^v - \sum_{\beta=1}^{t'} c^d_\beta$. If $F'$ does not hold, at least one true variable
part selects $x_i$; this variable part maps a heavy vertex of weight $c^v+c^u$
to the midpoint, while all other variable parts map a heavy vertex of weight at least $c^v$ to it.
Together with the floor vertex and vertices from the disjunction  parts, this gives a total weight
of the vertices mapped to the midpoint
that is at least $c - c^p +c^u > c$. We again have a contradiction, so $F'$ is satisfied.
\end{proof}

\begin{claim}
If element $F'$ is selected, then $F'$ is satisfied under the defined truth assignment.
\label{claim:demononstrated2true}
\end{claim}

\begin{proof}
We show this by induction.
Claim~\ref{claim:demonstratedliteral} shows
the result holds for literals. 

If $F'$ is a conjunction, then each term of $F'$ is selected (by definition), and thus satisfied (by induction).

If $F'$ is a disjunction, then by Claim~\ref{claim:selecteddisjunctionterm}, it has a
selected term, which is satisfied (by induction).
\end{proof}

As, by definition, $F$ is selected, we have that $F$ is satisfied by the defined truth assignment (by Claim~\ref{claim:demononstrated2true}). By 
Claim~\ref{claim:exactktrue}, this truth assignment sets exactly $k$ variables to true.
This ends the proof of Lemma~\ref{lemma:snake2formula}.
\end{proof}

\begin{lemma}
If $F$ can be satisfied by setting exactly $k$ variables to true, then there is a uniform emulation of
$P_N$ to $P_M$ with weight function $w$.
\label{lemma:formula2snake}
\end{lemma}

\begin{proof}
We describe how to build a uniform emulation of $P_N$ on $P_M$, given a truth assignment that satisfies
$F$ by setting exactly $k$ variables to true. Suppose the set of true variables is
$\{x_{t_1}, x_{t_2}, \ldots, x_{t_k}\}$.

\paragraph{Folding a path}
In the construction, we repeatedly use a simple step that `folds' a path of vertices of weight 1.
Suppose we have a vertex $x$ mapped to $i$ and a vertex $y$ mapped to $j$ with a path between $x$ and $y$
with $\gamma$ internal vertices.
Suppose $j\geq i$. We must have that $\gamma \geq j-i-1$, otherwise we cannot
find an emulation for the path from $i$ to $j$. To fold this path to the right, the $\delta$th vertex
on this path is mapped to $i+\delta$ when $\delta \leq (j-i) + \frac{\gamma - j+i}{2}$,
and to $j + (\gamma -\delta) +1$ otherwise. The illustration in Figure~\ref{figure:foldedpath} makes
the simple construction clear. We can also use a similar method to fold a path to the left. Note that
when we fold a path, each integer has at most two vertices of the folded path mapped to it.

\begin{figure}[htb]
    \centering
    \includegraphics{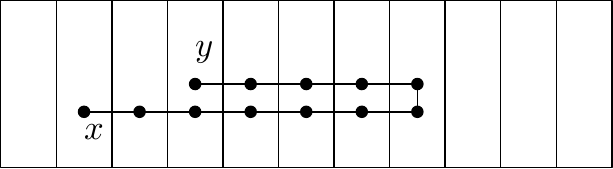}
    \caption{Folding a path from $x$ to $y$ to the right}
    \label{figure:foldedpath}
\end{figure}

We can now define the emulation of $P_N$ on $P_M$.

\paragraph{Floor} The $i$th vertex of the follow is mapped to $i$, $1\leq i \leq M$.

\paragraph{True variable parts} We map the vertices of $i$th true variable parts as follows:
\begin{itemize}
    \item The right turning point is mapped to $M$.
    \item The next path with $M-2$ vertices of weight 1 are mapped in order to $M-1$, $M-2$, \ldots, $2$.
    \item The left turning point is mapped to $1$.
    \item Recall that the $i$th variable that is set to true is $X_{t_i}$. We map the first heavy vertex
    of the $i$th true variable part to $t_i+2$, and fold the path of weight-one vertices between the left turning
    point and this first heavy vertex to the right.
    \item The heavy vertices are mapped to consecutive vertices, i.e., the $j$th heavy vertex
    is mapped to $j+t_i+1$. (Note that we thus set the $n$th heavy vertex to $n+1+t_i$.)
    \item We fold the path with weight-1 vertices between the last heavy vertex and the right turning point
    of the next part (the $i+1$st true variable part, or, if $i=k$, the first disjunction  part)
    to the left.
\end{itemize}

\paragraph{Disjunction parts}
We map the vertices of the $\alpha$th disjunction part as follows:
\begin{itemize}
    \item The right turning point is mapped to $M$.
    \item The next path with $M-2$ vertices of weight 1 are mapped in order to $M-1$, $M-2$, \ldots, $2$.
    \item The left turning point is mapped to $1$.
    \item We map each $i$th left anchor to the corresponding $i$th left anchor point of disjunctions at
    this or-depth. 
    \item For each disjunction $F'$, let $F''$ be a satisfied term of $F'$, if $F'$ is satisfied (by the
  given  truth assignment), otherwise, let $F''$ be the first term of $F'$. (We can take any arbitrary term in this case.) Now, for $1\leq i\leq 3m(F')-1$, place the $i$th selecting vertex of $F'$ at
  $LA(F'')-m+i$. 
    \item All paths of vertices of weight one (between turning points, anchors, first heavy vertices
    for disjunctions of this depth, and last heavy vertices for disjunctions of this depth) are folded.
    Most of these are folded to the right, but where necessary (as we cannot map vertices to $M$ or integers larger than $M$) we fold to the left. The last path of vertices of weight 1 of the
    last disjunction  part is folded (to the left), such that its last vertex is mapped to $M-1$.
\end{itemize}

\paragraph{Filler path}
Now, define, for $1\leq j \leq M$, $z_j$ to be the total weight of all floor, true variable parts, and
disjunction  parts that are mapped to $j$. We will later show that for all $j$, $z_j \leq c$,
and $z_1=z_M=c$ (see Claims~\ref{claim:z1}, \ref{claim:z2}, \ref{claim:z3}, and \ref{claim:z4}.)

Note that in the construction above,
the predecessor of the first vertex on the filler path is mapped to $M-1$: this is the last vertex
of the last disjunction  part.

We can now use the filler path to ensure that each vertex on $P_M$ has a total weight $c$ mapped to it.
Repeat the following step, for $j$ from $M-1$ to $2$: place the next $c-z_j$ vertices of the filler path at $j$.

\begin{claim}
$z_1 = z_M = c$.
\label{claim:z1}
\end{claim}

\begin{proof}
To $1$, we map the first floor vertex, and $k+t'$ left turning points; the total weight of these vertices equals $c$. To $M$, we map the last floor vertex, and $k+t'$ right turning points; the total weight of these vertices equals $c$. No other vertices are mapped to $1$ or $M$.
\end{proof}

\begin{claim}
Each integer in $[2,M-1]$ has less than $c^p$ vertices of weight one from true variable parts and
disjunction  parts mapped to it.
\label{claim:zweigthone}
\end{claim}

\begin{proof}
Let $i\in [2,M-1]$. From each true variable part, there are at most three vertices of weight 1 mapped to $i$: one vertex
from the first path with $M-2$ vertices, and at most two vertices from one of the two folded paths of
length $n-1$.
From each disjunction  part, there are at most five vertices of weight one mapped to $i$:
again one vertex from the first path with $M-2$ vertices, and at most four vertices from folded
paths (possibly, one path folded to the left, and one path folded to the right; each folded path contributes at most two vertices that are mapped to $i$).
\end{proof}

\begin{claim}
Let $i\in [2,2n+1]$. Then $z_i < c$.
\label{claim:z2}
\end{claim}

\begin{proof}
To $i$, we mapped (apart from filler vertices), one floor vertex, at most $c^p$ vertices of weight one
(Claim~\ref{claim:zweigthone}), and at most $k$ heavy vertices of true variable parts.
When $2\leq i \leq n+1$, each of these heavy vertices of true variable parts has weight $c^v$;
when $n+2\leq i \leq 2n+1$, at most one of these heavy vertices of true variable parts has weight
$c^v+c^u$: the weight of the $j$th heavy vertex equals $c^v+c^u$, if and only if
the $j$th variable that is set to true is $x_{i-n-1}$ --- as we set exactly $k$ variables to true,
these are all different for different $j\in \{1, \ldots, k\}$.
No other vertices are mapped to $i$.
\end{proof}

\begin{claim}
Let $i\in [M-n-1,M-1]$. Then $z_i < c$.
\label{claim:z3}
\end{claim}

\begin{proof}
This case is similar to the case that $i \in [2,n+1]$; see the proof of Claim~\ref{claim:z2}.
\end{proof}

\begin{claim}
Let $i\in [2n+2,M-n-2]$. Then $z_i < c$.
\label{claim:z4}
\end{claim}

\begin{proof}
We have less than $c^p$ vertices of weight one mapped to $i$ (Claim~\ref{claim:zweigthone}),
$k$ heavy vertices of true variable parts, and a floor vertex.

If we map an anchor vertex to $i$, then $i$ is the corresponding anchor point, and we subtracted the weight of this anchor vertex from the weight of the floor vertex that is mapped to $i$.

If we map a heavy vertex of a disjunction of or-depth $\alpha$, then $i$ is between the anchor points
of this disjunction, and we subtracted the weight of this heavy vertex from the weight of the floor vertex mapped to $i$.

We now consider the additive terms $c^u$, subtracted from the weight of the floor vertex mapped to $i$,
and added to the weight of heavy vertices of true variable parts. There are a number of cases.

Case 1: $i$ is the midpoint of the interval of a positive literal that is satisfied. We subtract
$(k-1)\cdot c^u$ from the weight of the $i$th floor vertex, and at most $k-1$ heavy vertices of
true variables parts have weight $c^v+c^u$: if the $j$th variable set to true satisfies the literal,
then by construction, the $j$th true variable part has a heavy vertex of weight $c^v$ mapped to it.

Case 2: $i$ is the midpoint of the interval of a negative literal, say $\neg x_j$ that is satisfied. 
So, $x_j$ is not selected to be true, and by construction, all true variable parts have a
vertex of weight $c^v$ mapped to $i$.

Case 3: $i$ is the midpoint of the interval of a literal $\ell$ that is not satisfied. We can have
$k$ heavy vertices of true variable parts with weight $c^v+c^u$ mapped to $i$. 
Let $q$ be the element that is an ancestor of $\ell$ or $\ell$ itself, that is not satisfied,
and that has minimum or-depth among all such elements. $q$ cannot be $F$ as $F$ is satisfied.
The parent of $q$ is satisfied, thus must be a disjunction. 
Now, the selecting vertices of this disjunction are mapped to a range
that is disjoint from the interval of $\ell$. We have a term $c^d_\alpha$ subtracted from the
weight of the floor vertex mapped to $i$, but no heavy vertex of a disjunction  part
of weight $c^d_\alpha$ mapped to $i$. Now note that $c^d_\alpha > k \cdot c^u$.

Case 4: $i$ is not the midpoint of the interval of a literal. At most $k$ heavy vertices
of true variable parts mapped to $i$ have weight $c^v+c^u$, and we subtracted $k \cdot c^u$ from the
weight of the $i$th floor vertex.

In each of the cases, we have a term that is subtracted from the weight of the floor vertex mapped to $i$
that is at least the total of all additive terms $c^u$ of heavy vertices of true variable parts that are mapped to $i$. The claim now follows.
\end{proof}

It is not hard to see that the mapping we defined is indeed an emulation. The filler path ensures that
the total weight mapped to each vertex equals $c$; Claims~\ref{claim:z1}, \ref{claim:z2}, \ref{claim:z3}, and \ref{claim:z4} show that the mapping of the filler path indeed is possible. This finishes the proof
of Lemma~\ref{lemma:formula2snake}.
\end{proof}

Lemmas~\ref{lemma:snake2formula} and \ref{lemma:formula2snake} together give
the main result of this section.

\begin{theorem}
{\sc Weighted Path Emulation} is $W[t]$-hard for all positive integers $t$.
\label{theorem:wpe}
\end{theorem}

We need in the next section actually a slightly different result (for an easier proof), namely, we require that the first vertex of $P_N$ is mapped to $M$. 
From the proof above, it is clear that we can require this or require that $f(1)=1$. (The latter
is imminent from the proof; the former follows by mirror the map $f$ and use $g(v) = M+1-f(v)$.)

\begin{corollary}
{\sc Weighted Path Emulation with $f(1)=1$} 
and {\sc Weighted Path Emulation with $f(1)=M$} are $W[t]$-hard for all positive integers $t$.
\label{cor:uep1}
\end{corollary}

For the hardness proof of {\sc Directed Bandwidth}, we need a small variation of the proof.

\begin{corollary}
{\sc Weighted Path Emulation with $f(1)=1$ and $f(N)=1$} 
and {\sc Weighted Path Emulation with $f(1)=M$ and $f(N)=M$} are $W[t]$-hard for all positive integers $t$.
\end{corollary}

\begin{proof}
We can use the same proof as for Theorem~\ref{theorem:wpe}, with the following modification: we subtract one from the weight of the first floor vertex, and make the filler path one longer. Now, if there is
a uniform emulation, there is one where the first floor vertex and the last vertex of the filler path are mapped to 1; the second result is again obtained by mirroring the map.
\end{proof}

A direct consequence of our proof is that {\sc Weighted Path Emulation} is NP-complete, even when all weights are given in unary.

\begin{corollary}
{\sc Weighted Path Emulation} is strongly NP-complete. 
\end{corollary}

\begin{proof}
We use a transformation from {\sc 3-Satisfiability}. Suppose we have a formula $F$ with $n$ variables.
Add $n$ additional variables that do not appear in $F$, and ask if we can satisfy $F$ by setting
exactly $n$ out of the $2n$ variables to true. Now, use the transformation above. 

Observe that, as $t'=1$, we have that $N$, $M$, and all vertex weights are polynomially bounded in $n$.
As membership in NP is trivial, the result now follows again from Lemmas~\ref{lemma:snake2formula} and \ref{lemma:formula2snake}.
\end{proof}

\section{Bandwidth of caterpillars}
\label{section:bandwidth}

In this section, we show the following theorem:

\begin{theorem}
{\sc Bandwidth} is $W[t]$-hard for all positive integers $t$, when restricted to caterpillars with hair length at most three.
\label{theorem:bandwidth}
\end{theorem}

To prove Theorem~\ref{theorem:bandwidth}, we transform from the {\sc Weighted Path Emulation with $f(1)=M$} problem.

Suppose we are given paths $P_N$ and $P_M$, with a weight function $w: \{1, \ldots, N\} \rightarrow \{1, \ldots, c\}$, and suppose $c = \sum_{i=1}^N w(i) / M$ is the emulation factor.
Thus, $\sum_{i=1}^N w(i) = cM$. Recall that we parameterized this problem by the value $c$.

Assume that $c> 6$; otherwise, obtain an equivalent instance by multiplying all weights by 7.

Let $b= 12c+6$. Let $k= 9bc +b$. Note that $k$ is even.
We give a caterpillar $G=(V,E)$ with hair length at most three, with the property that
$P_N$ has a uniform emulation on $P_M$, if and only if $G$ has bandwidth at most $k$.

\begin{figure}
    \centering
    \includegraphics[width=.95\linewidth]{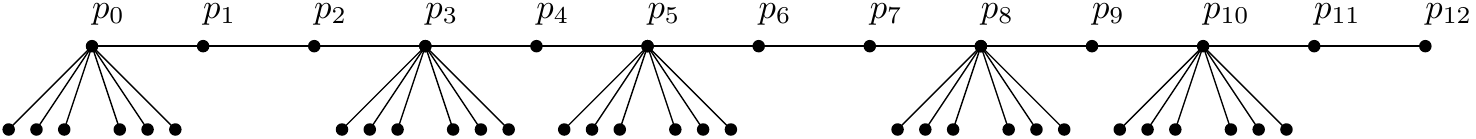}
    \caption{First part of the caterpillar}
    \label{fig:spinepath}
\end{figure}

\begin{figure}[htb]
    \centering
    \includegraphics{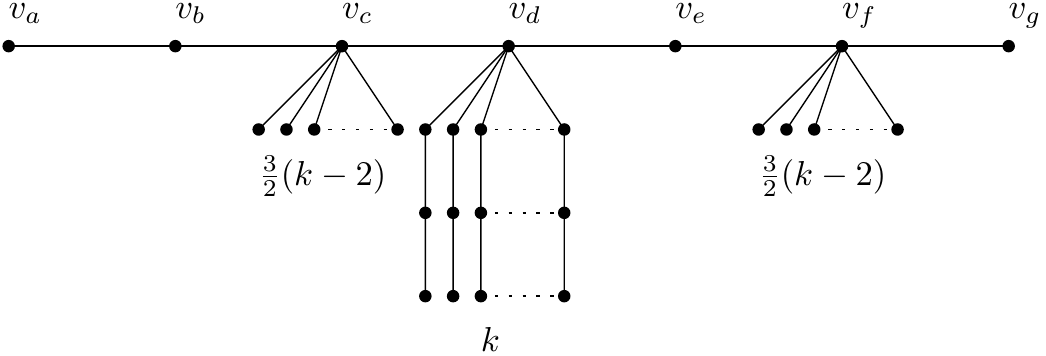}
    \caption{The Turning Point, after Monien~\cite{Monien86}}
    \label{fig:turningpoint}
\end{figure}

$G$ is constructed in the following way:

\begin{itemize}
    \item We have a {\em left barrier}: a vertex $p_0$ which has $2k-1$ hairs of length one, and is neighbor to $p_1$.
    \item We have a path with $5M-3$ vertices, $p_1, \ldots, p_{5M-3}$. As written above, $p_1$ is 
    adjacent to $p_0$. Each vertex of the form $5i-2$ or $5i$ ($1\leq i\leq M-1$) receives
    $2k-2b$ hairs of length one. See Figure~\ref{fig:spinepath}. We call this part the {\em floor}.
    \item Adjacent to vertex $p_{5M-3}$, we add the {\em turning point} from the proof of Monien~\cite{Monien86}. 
    We have vertices $v_a = p_{5M-3}$, $v_b$, $v_c$, $v_d$, $v_e$, $v_f$, $v_g$, which are successive vertices on
    a path. 
    I.e., we identify one vertex of the turning point ($v_a$) with the last vertex of the floor $p_{5M-3}$.
        To $v_c$, we add $\frac{3}{2} (k-2)$ hairs of length one; to $v_d$, we add $k$ hairs of length three, and to $v_f$ we add $\frac{3}{2} (k-2)$ hairs of length one. Note that this construction is identical to the one by Monien~\cite{Monien86}; vertex names are chosen to facilitate comparison with Moniens proof. See Figure~\ref{fig:turningpoint}.
    \item Add a path with $6N-5$ vertices, say $y_1, \ldots y_{6n-5}$, with $y_1$ adjacent to $v_g$. To each vertex of the form $y_{6i-5}$, add $ 9b \cdot w(i)$ hairs of length one.
    We call this part the {\em weighted path gadget}.
    \item Note that the number of vertices that we defined so far and that is not part of the turning point equals
    $2k + 5M-3 + 2(M-1)(2k-2b)+ 6N-5 + 9b \sum_{i=1}^N w_i = 5M + 4Mk - 2k - 4Mb + 4b + 9bcM$.
    Let this number be $\alpha$. One easily sees that $\alpha \leq (5M-2)k -1$.
   Add a path with $(5M-2)k-1-\alpha$ vertices and make it adjacent to $y_{6n}$.
    We call this the {\em filler} path.
\end{itemize}

Let $G$ be the remaining graph. Clearly, $G$ is a caterpillar with hair length at most three. It is interesting to note that the hair lengths larger than one are only used for the turning point.

\begin{figure}[htb]
    \centering
    \includegraphics[width= \textwidth]{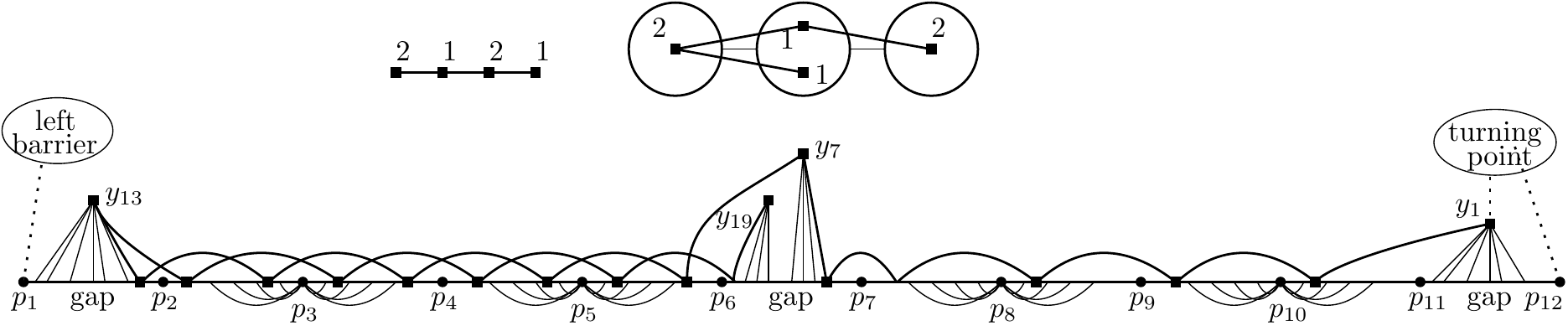}
    \caption{Illustration of part of the construction. Shown are $P_4$ with successive vertex weights 2, 1, 2, 1; a uniform emulation on $P_3$ with emulation factor 2; a layout of a part of $G$.}
    \label{figure:examplebw}
\end{figure}

\paragraph{Some intuition}
The main ideas behind the proof are the following. The construction is such that $p_0$ with hairs
is at one side of the layout, and the turning point is at the other side of the layout. The total
number of vertices enforces that successive vertices $p_i$, $p_{i+1}$ always have distance exactly $k$.
The hairs of vertices $p_3$, $p_5$, $p_8$, etc.\ fill `most of' the spots in the neighboring intervals;
this gives `gaps' between $p_1$ and $p_2$, between $p_6$ and $p_7$, etc. (See Figure~\ref{fig:spinepath}.)
Vertices of the form 
$y_{6i-5}$ with (most of) their hairs only fit in these gaps; if $y_{6i-5}$ is placed in the
gap between $p_{5j-4}$ and $p_{5j-3}$, then we set $f(i)=j$. Because $y_{6i-5}$ and $y_{6i+1}$ have distance six in $G$, they must be placed in the same or neighboring gaps, and thus $|f(i)-f(i+1)|\leq 1$, i.e., $f$ is an emulation. The number of hairs of vertices $y_{6i-5}$ and sizes of gaps is such that
the total weight of vertices in one gap is bounded by $c$, and thus it follows that $f$ is uniform.
See also Figure~\ref{figure:examplebw}.

We now give the formal correctness proof of our construction.

\begin{lemma}
Suppose $P_N$ has a uniform emulation $g$ on $P_M$ with emulation factor $c$ with $f(1)=M$. Then the bandwidth of $G$ is at most $k$.
\label{lemma:ue2bw}
\end{lemma}

\begin{proof}
Suppose we are given a uniform emulation $f: \{1, \ldots, N \} \rightarrow \{1, \ldots, M\}$ of $P_N$ on
$P_M$ (for weight function $w$). We now can construct a layout $g$ 
of $G$ with bandwidth at most $k$ as follows.

For $1 \leq i \leq 5M-3$, set $g(p_i) = (i+1)k+1$.
Each of the hairs of $p_0$ is mapped by $g$ to
a different value in $\{1, \ldots, 2k\} \setminus \{k+1\}$. (We later define how to map the hairs of
other vertices of the floor.)

For the turning point, we set $g(v_g)=g(v_a)-1 = (5M-2)k$. The other vertices of the turning point
receive values larger than $(5M-2)k$; the remaining layout for the turning point is identical to the one described by Monien~\cite{Monien86}.

Each position of the form $(i+1)k+1 + j$ for $1\leq i\leq 5M-3$ with $j\in \{-1,+1\}$ is said
to be {\em reserved}. Note these are the positions before and after the image of the vertices $p_i$.
These positions are reserved for the filler path.

We call the positions between $g(p_{5i-4})$ and $g(p_{5i-3})$ the $i$th {\em gap}, for $1\leq i\leq M$.
Gaps $1$ till $M-1$ have at this point $k-1$ unused integers of which two are reversed; in
gap $M$ we have $k-1$ unused integers ($v_g$ is mapped in the gap) of which one is reserved.
(The gaps are used to layout `most of' the vertices of the weighted path gadget.)

We now show how to map the weighted path gadget. We start with fixing the images of vertices
of the form $y_{6i-5}$.
For each $i$, $1\leq i \leq N$, set $g(y_{6i-5})$ to be the first integer in the $f(i)$th gap 
that is so far not used and
is not reserved. As there are at most $c$ vertices of $P_N$ mapped by $f$ to a specific vertex of $P_M$,
we have that $g(y_{6i-5})$ is in the interval $[f(p_{5(i-1)})+3, f(p_{5(i-1)})+c+2]$,
thus $|g(y_{6i-5}) - g(y_{6i+1}) | \leq 5k+c$; i.e., they can be in the same, or neighboring gaps.

As $f(1)=M$, we have that $y_1$ is placed in the $M$th gap, as is, by construction its neighbor
$v_g$; thus $|g(y_1)-g(v_g)|\leq k$.

The next step is to map the vertices of the form $y_j$ with $j$ not of the form $6i-5$. We must do this
in such a way, that neighboring vertices are mapped to integers that differ at most $k$, taking into
account the mapping vertices of the form $y_{6i-5}$ as defined above. 

So, we consider, for $i \in \{1, \ldots, N-1\}$, the vertices on the path from $y_{6i-5}$ to $y_{6i+1}$ are mapped as follows.
If $y_{6i-5}$ and $y_{6i+1}$ are mapped to the same gap, then map the five vertices on this path also to this gap, to so far unused and unreserved values. 
Otherwise, define a `largest possible step to the right' from an integer $\alpha$ as the largest integer
that is at most $\alpha+k$ and is not used or reserved. If $g(y_{6i-5}) < g(y_{6i+1})$, then put
$g(y_{6i-4})$ at the largest possible step to the right from $g(y_{6i-5})$, and then 
$g(y_{6i-3})$ at the largest possible step to the right from $g(y_{6i-4})$, and continue doing so,
till we placed $g(y_{6i})$.  If $g(y_{6i+1}) < g(y_{6i-5})$, a similar construction is followed. 

\begin{claim}
For each $j$, $1\leq j \leq 5M-4$, we map at most $11c$ vertices of the form $y_i$ to integers in
the interval $[g(p_j), g(p_{j+1})]$.
\label{claim:used}
\end{claim}

\begin{proof}
If the interval is a gap, then for each such $y_i$, there is a vertex of the form $y_{6'-5}$ s at distance at most five from $y_i$ mapped to this gap. As we have at most 
$c$ vertices of the form $y_{6i'-5}$ in the gap, we have at most $11c$ vertices of the form $y_i$ mapped to this gap.

If the interval is not a gap, then for each such $y_i$, there is a vertex of the form $y_{6'-5}$ s at distance at most five from $y_i$ mapped to the last gap before the interval. Thus, we have
at most $10c$ vertices of the form $y_i$ mapped to this gap. (A more precise counting argument can give a smaller bound.)
\end{proof} 

\begin{claim}
For all $i$, $1\leq i < N$, $|g(y_{6i}) - g(y_{6i+1}| \leq k$.
\label{claim:4.4}
\end{claim}

\begin{proof}
The result clearly holds when $y_{6i-5}$ and $y_{6i+1}$ are placed in the same gap. If they are
placed in different gaps, then we consider the
case that $g(y_{6i-5}) < g(y_{6i+1})$; the other case is similar.
 
Note that each largest possible step to the right makes a jump that is at least $k - 22c-5$: from
any $k$ consecutive integers, we have used at most $22c$ for vertices of the form $y_i$ (by Claim~\ref{claim:used}, with vertices possible in two consecutive intervals), reserved four integers,
and used one for a vertex of the form $p_i$. 

Recall that $|g(y_{6i-5}) - g(y_{6i+1}) | \leq 5k+c$. Five largest jumps to the right bridge
at least $5 (k- 22c - 5)$, hence $g(6i+1)-g(6i) \leq 23c+25 < k$.
\end{proof}

The next step is to map an initial part of the filler part, to `move it' to the last number in the $M-1$st
gap. 
Suppose $f(v_N)=i$. Now, map the first vertex of the filler
path to $f(p_{5i-3})-1$, i.e., the last number in the $i$th gap (which was reserved), and then,
while we did not yet place a vertex on $f(p(5(M-1)-3))-1$ (the last number in the $M-1$st gap), map
$k$ larger than its predecessor on the path,
i.e., the second vertex goes to $g(p_{5i-3})+k-1$, the third to $g(p_{5i-3})+2k-1$, etc. In each
case, we place the vertex on the last number in the gap; these were reserved so not used by other vertices.

At this point, each open interval $(g(p_i), g(p_{i+1})$ has at most $11c+1$ used positions, and at at most
four reserved positions. Thus, at least $k-1-(11c+5) \geq k-b = 9bc$ positions are unused and not reserved
in this interval. 

For each vertex of the form $p_{5i-2}$ or $p_{5i}$ $(1\leq i \leq M-1)$, 
say $p_{i'}$
we map $k-b$ of its hairs to the interval before it: i.e., to integers in $(g(p_{i'-1}), g(p_{i'}))$ that
are unused and not reserved, and $k-b$ hairs to the interval after it, i.e., to integers in $(g(p_{i'}), g(p_{i'+1}))$ that
are unused and not reserved. 

\begin{claim}
For each $j$, $1\leq j\leq M$, the total number of hairs adjacent to vertices of the form $y_{6i-5}$, $1\leq i\leq N$ with $y_{6i-5}$ mapped to the $j$th gap equals $9bc$.
\label{claim:hairsingap}
\end{claim}

\begin{proof}
As $f$ is uniform, we have that $\sum_{i: f(i)=j} w_i = c$. A vertex $y_{6i-5}$ is placed in the $j$th gap, if and only if it contributes to this sum. As the number of hairs of $y_{6i-5}$ equals
$9b\cdot w(i)$, the result follows.
\end{proof}

For each vertex of the form $y_{6i-5}$, $1\leq i\leq N$, we map all its hairs to the gap to which it
belongs. I.e., if $f(i)=j$, then all hairs of $y_{6i-5}$ are mapped to unused and not reserved
integers in $(g(p_{5j-4},g(p_{5j-3})$. From the analysis above, we have sufficiently many such integers
to map all hairs.

Finally, we lay out the filler path to fill up all remaining unused values: go from $f(p(5(M-1)-4))-1$
to the last unused number of the $M$th gap (which is smaller than $f(p(5(M-1)-4))+k-1$, and then,
while there are unused numbers left, place each next vertex of the filler gap on the largest so far unused number.
Thus, from this point on, the filler gap goes from right to left, skipping only used numbers. 
The reserved numbers of the form $(i+1)k+1$ guarantee that by doing so, we never skip more than $k-1$ numbers, thus preserving the bandwidth condition.

This finishes the construction of the linear ordering $g$. From the analysis above, it follows that $g$
has bandwidth $k$.
\end{proof}

\begin{lemma}
Suppose the bandwidth of $G$ is at most $k$. Then $P_N$ has a uniform emulation on $P_M$ with emulation factor $c$ with $f(1)=M$.
\label{lemma:uetobw}
\end{lemma}

\begin{proof}
Suppose we have a linear ordering $g$ of $G$ of bandwidth at most $k$.

First we note that all vertices outside the left barrier and the turning point have to be mapped between
the left barrier and the turning point. For the left barrier, this is easy to see: $p_0$ has $2k$ neighbors, which will occupy all $2k$ positions with distance at most $k$ to $g(p_0)$, and thus all other vertices
must be at the same side of $p_0$ as $p_1$. 

Now, we look at the turning point, and use a result by Monien~\cite[Lemma 1]{Monien86}.

\begin{lemma}[Monien~\cite{Monien86}]
Let $g$ be a linear ordering of a graph $G$ containing the turning point as subgraph.
$|g(v_a)-g(v_g)|\leq 1$, and one of the following two cases holds.
\begin{enumerate}
    \item All vertices in the turning point except $v_a$ and $v_g$ have an image under $g$ that is
    larger than $\max\{g(v_a), g(v_g)\}$; all vertices in $G$ that do not belong to the turning point have an image that is smaller than $\min\{g(v_a), g(v_g)\}$.
        \item All vertices in the turning point except $v_a$ and $v_g$ have an image under $g$ that is
    smaller than $\min\{g(v_a), g(v_g)\}$; all vertices in $G$ that do not belong to the turning point have an image that is larger than $\max\{g(v_a), g(v_g)\}$.
\end{enumerate}
\end{lemma}

I.e., we have that $v_a$ and $v_g$ are next to each other; all other vertices of the
turning point are at one side of this pair, and all vertices not in the turning point at the other side.
Without loss of generality, suppose $g(p_0) < g(v_a)$. 

Thus, from left to right we have:
$k$ hairs of $p_0$; $p_0$; $k-1$ hairs of $p_0$; $p_1$; all vertices except $p_1$, vertices in the left barrier, and vertices in the turning point; $v_a = p_{5M-3}$ and $v_g$ (in this order, or reversed); all other vertices of the turning point.

\begin{claim}
$g(p_0)=k+1$, and $\{g(v_a),g(v_g)\} = \{(5M-2)k, (5K-2)k+1\}$.
\end{claim}

\begin{proof}
$p_0$ is placed after $k$ hairs, so must have image $k+1$.
The total number of vertices not in the turning point equals $(5M-2)k-1$, see the construction
of the filler path. The pair $\{g(v_a),g(v_g)\}$ comes after all vertices not in the turning point,
and before all other vertices from the turning point, so must go to $(5M-2)k$ and $(5M-2)k+1$.
\end{proof}

\begin{claim}
For all $i$, $0\leq i < 5M-3$, 
either $g(p_{i+1}) = g(p_i) +k$ or $g(p_{i+1}) = g(p_i) +k-1$. 
\end{claim}

\begin{proof}
The path from $p_0$ to $v_a=p_{5M-3}$ has $M-3$ edges. For each pair $p_i$, $p_{i+1}$, the difference
of the images is at most $k$ (by the bandwidth condition), but the total of the differences over all pairs on the path must
be at least $(M-3)\cdot k -1$.
\end{proof}

Now, call the positions between $g(p_{5i-5})$ AND $g(p_{5i-2})$ the $i$th {\em enlarged gap}, for $1\leq i \leq M-1$; and between $g(5M-5)$ and $g(p_{5M-3})$ the $M$th enlarged gap. 

\begin{claim}
Each vertex of the form $y_{6j-5}$ is mapped to an enlarged gap.
\label{claim:gap1}
\end{claim}

\begin{proof}
Suppose not. As $g(y_{6j-5}) > g(p_1)$ and $g(y_{6j-5}) < g_{5M-3}$, by construction of the left barrier
and turning point, we have that there is an $i$ with $g(y_{6j-5}) \in [g(p_{5i-2}), g(p_{5i})]$. 
Now, $[g(p_{5i-3}), g(p_{5i+1})]$ contains $4k-4b$ hairs attached to vertices $p_{5i-2}$ and $p_{5i}$
and $9b\cdot w(j) \geq 9b$ hairs attached to $y_{6j-5}$. The interval has size at most $4k$, but has
$4k+b$ hairs mapped to it; contradiction.
\end{proof}

Now, let $f:\{1,\ldots, N\} \rightarrow \{1, \ldots, M\}$, such that $f(j)=i$, if $y_{6j-5}$
is mapped to the $i$th enlarged gap. By Claim~\ref{claim:gap1}, $f$ is well defined.

\begin{claim}
$f$ is an emulation.
\label{claim:gap2}
\end{claim}

\begin{proof}
Note that there is a path with six edges from a vertex $y_{6j-5}$ to $y_{6(j+1)-5}$. As the distance
between vertices $p_i$ in the linear ordering is either $k$ or $k-1$, each edge in this path can
jump over at most one vertex of the form $p_\alpha$. To go from the $i$th enlarged gap to the $i+2$nd enlarged gap,
we must jump at least seven vertices of the form $p_\alpha$, so $y_{6j-5}$ and $y_{6(j+1)-5}$
are mapped to the same or neighboring enlarged gaps, thus $|f(j)-f(j+1)|\leq 1$.
\end{proof}

\begin{claim}
$f$ is uniform.
\label{claim:gap3}
\end{claim}

\begin{proof}
We show that for each $i$, $\sum_{j: f(j)=i} w(j) \leq c$. As $f$ is uniform, we have
that for each $i$: $\sum_{j: f(j)=i} w(j) = c$. 

We only give the proof for $1 < i < M$; the cases $i=1$ and $i=M$ are similar (using that no vertices
can be mapped to values used by the left blockade and turning point) and omitted.

Consider an $i$, $1<i<M$. 
For each $j$ with $f(i)=j$, all $9b \cdot w(j)$ hairs of $y_{6j-5}$ are mapped to the interval
$[g(p_{5i-6}), g(p_{5i-1})]$. This interval has size at most $5k$, contains $4k-4b$ hairs of
$p_\alpha$-vertices, thus can contain at most $8k+4b$ hairs of vertices $y_\beta$. All hairs of
vertices $y_{6j-5}$ with $f(j)=i$ must be mapped to this interval. As the number of these hairs
equals $9b$ times the sum of the weights of these vertices, we have $\sum_{j: f(j)=i} w(j) \leq c$.
\end{proof}

Finally, we observe that $f(1)=M$: $y_1$ is incident to $v_g$, and thus $y_1$ must be placed in
the $M$th enlarged gap.

Now, with this last observation, Claims~\ref{claim:gap1}, \ref{claim:gap2} and \ref{claim:gap3} together show Lemma~\ref{lemma:uetobw}.
\end{proof}

As the construction of the caterpillar $G$ can be done in polynomial time, given $M$, $N$ and $w$,
the main result of this section now follows.

\begin{theorem}
{\sc Bandwidth} for caterpillars with hair length at most three is $W[t]$-hard for all $t\in {\bf N}$.
\end{theorem}

\section{Directed Bandwidth}
\label{section:directedbandwidth}
A minor variation of the proof of Theorem~\ref{theorem:bandwidth} gives the following result.

\begin{theorem}
{\sc Directed Bandwidth} is hard for $W[t]$ for all positive integers $t$, when restricted to
directed acyclic graphs whose underlying undirected graph is a caterpillar with hair length at most one.
\label{theorem:directedbandwidth}
\end{theorem}

The proof of Theorem~\ref{theorem:directedbandwidth} is similar to the proof of Theorem~\ref{theorem:bandwidth}. Instead of repeating many details, we list the differences.

The first difference is that we reduce from {\sc Weighted Path Emulation with $f(1)=M$ and $f(N)=M$.}

\paragraph{Constants}
Instead of setting $b=12c+6$, we take $b=24c+6$. We still set $k=9bc+b$.

\paragraph{The right turning point}
A simpler construction can be used to the right turning point. We add $k$ `right turning point' vertices, each with
indegree 1 and outdegree 0, with an arc from $p_{5M-3}$ to each of these.

Note that each of these vertices must be mapped to the right of $p_{5M-3}$. The $k$ positions right
of the image of $p_{5M-3}$ are thus used by these right turning point vertices, and thus all other vertices must be mapped to positions left of $p_{5M-3}$.

\paragraph{Directing edges}
$k$ of the hairs of $p_0$ are directed to $p_0$, and the other $k-1$ hairs of $p_0$ are directed out of
$p_0$. The other hairs of floor vertices are directed as follows: each vertex of
the form $p_{5i-2}$ or $p_{5i}$ has $k-b$ hairs directed to it, and $k-b$ hairs directed out of it.

All edges $\{p_i,p_{i+1}\}$ are changed to an arc $(p_i,p_{i+1})$. 

All hairs in the weighted path gadget are directed from vertex on the spine (of the form $y_{6i-1})$) to
the hair vertex, i.e., the spine vertex is tail of the arc.

We start the filler path with an edge from the first to the second vertex. All other vertices on
the filler path are directed in the reverse direction, i.e., for $i\geq 2$, the edge between
the $i$th and $i+1$st vertex from the filler path is directed from the $i+1$st vertex to the $i$th vertex.

\begin{figure}
    \centering
    \includegraphics{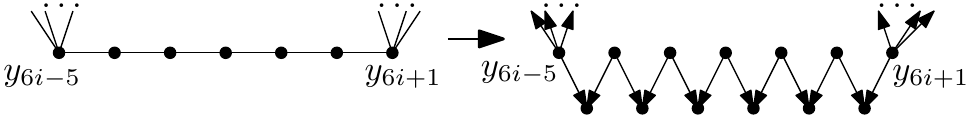}
    \caption{Change of construction for the weighted path gadget}
    \label{figure:directedbwgadget}
\end{figure}

\paragraph{Paths between vertices in the weighted path gadget}
Each edge $\{y_i,y_{i+1}\}$ in the weighted path gadget is replaced by two arcs to a new vertex
$z_i$: we take arcs $(y_i,z_i)$ and $(y_{i+1},z_i)$. Figure~\ref{figure:directedbwgadget} illustrates
the change. 

\smallskip

Small modifications to the construction and proof are needed.

First, when we have a uniform emulation on $P_M$ with emulation factor $c$ and $f(1)=f(N)=M$,
then we can use a construction similar to Lemma~\ref{lemma:ue2bw}, with the following differences.

In addition to a `largest possible step' to the right (left), we also define a `smallest possible step to the right' (left) from an integer $\alpha$.
which is the smallest (largest) unused and unreserved integer larger (smaller) than $\alpha$. To layout
a path between two vertices $y_{6i-5}$ and $y_{6i+1}$, with $g(y_{6i+1}$ in the gap after $g(y_{6i-5})$.
we alternate a largest possible step to the right with a smallest possible step to the left. There is one exception: each vertex of the form $y_{6i+1}$ is always placed in the leftmost unused and unreserved integer in the gap where it has to be mapped to.
In the other cases, similar constructions are followed. A counterpart of Claim~\ref{claim:used} can be
shown, where the value $11c$ is replaced by $22c$. With help of the larger value of $b$, we again have
that Claim~\ref{claim:4.4} holds in this construction.

As we assume that $f(N)=M$, we have that $y_{6N-5}$ is mapped to the $M$th gap. We thus can start the
layout of the filler path by placing the first vertex of the filler path to the last reserved integer.
Then, the remainder of the filler path is mapped by always going to the highest not yet used integer.

The remainder is very similar to the proof of Lemma~\ref{lemma:ue2bw} and omitted here.

Second, when the bandwidth is $G$, a proof similar to the proof of Lemma~\ref{lemma:uetobw} shows that
there is a uniform emulation on $P_M$ with emulation factor $c$ and $f(1)=f(N)=M$. (E.g., see Figure~\ref{figure:directedbwgadget}; $g(y_{6i+5}) - g(y_{6i-1}) \leq 6k$, as there are six arcs where we
make a step of at most $k$ size to the right; the other arcs step to the left.)
Again, we omit most details here.

Finally, notice that the current construction gives a directed acyclic graph whose underlying undirected graph is a caterpillar with hair length at most one.

\section{Conclusions}
\label{section:conclusions}
In this paper, we showed that {\sc Bandwidth} is hard for the complexity class $W[t]$ for all positive
integers $t\in N$, even when the input graph is a caterpillar with hairs of length at most three. The proof uses some techniques and gadgets from the NP-completeness proof of {\sc Bandwidth} for this class of graphs 
by Monien~\cite{Monien86}. Monien also shows NP-completeness of {\sc Bandwidth} for caterpillars of maximum degree three (with arbitrary hair length). This raises the question whether {\sc Bandwidth} for caterpillars
with maximum degree three is $W[t]$-hard for all $t$. We conjecture that this is the case; perhaps with
a modification of our proof such a result can be achieved?

An intermediate result of independent interest is the $W[t]$-hardness of {\sc Weighted Path Emulation}. 
We used this result as a stepping stone for our main result, but expect that the result may also be useful for proving hardness for other problems as well.


It is unlikely that {\sc Bandwidth} belongs to $W[P]$.
In \cite{FellowsR20}, Fellows and Rosamond describe an argument, due to Hallett, that gives the intuition behind the conjecture that
{\sc Bandwidth} does not belong to $W[P]$. From the works of 
Bodlaender et al.~\cite{BodlaenderDFH09} and Drucker~\cite{Drucker15}, it follows that problems that are AND-compositional do not have a polynomial kernel unless $NP \subseteq coNP/poly$. The intuition behind 
this methodology is that such a polynomial kernel for an AND-compositional problem would give an unlikely strong compression of information. While {\sc Bandwidth} is not in FPT, assuming $W[t] \not\subseteq FPT$, for some $t$, and thus has no kernel (of any size), it is AND-compositional. If {\sc Bandwidth} would
belong to $W[P]$, it would have a certificate of $O(k \log n)$ bits (namely, the indices of the variables that
are set to true), and it is unlikely that an AND-compositional problem has such a small certificate.
We thus can formulate the following conjecture, due to Hallett.

\begin{conjecture}[Hallett, see also \cite{FellowsR20}]
{\sc Bandwidth} does not belong to $W[P]$, unless $NP \subseteq coNP/poly$.
\end{conjecture}

Finally, we conjecture that with modifications of the techniques from this paper, it is possible to show for more problems hardness for the $W[t]$-classes.

\section*{Acknowledgements}
I thank Michael Fellows and Michael Hallett for discussions and earlier joint work on
the parameterized complexity of {\sc Bandwidth}, as reported in \cite{BodlaenderFH94}. Some techniques in this paper are based upon techniques, underlying the results reported in \cite{BodlaenderFH94}.
I thank Marieke van der Wegen for discussions.

\bibliographystyle{abbrv}
\bibliography{bw}

\end{document}